\documentclass[%
 reprint,
 amsmath,amssymb,
 aps,
aps,prl,twocolumn,superscriptaddress,floatfix,longbibliography,showpacs]{revtex4-1}

\usepackage{graphicx}
\usepackage{dcolumn}
\usepackage{bm}
\usepackage{xcolor}
\usepackage{capt-of}
\usepackage{soul}
\usepackage{hyperref}
\hypersetup{
    colorlinks=true,
    allcolors=blue,
    pdftitle={Overleaf Example},
    pdfpagemode=FullScreen,
    }
\makeatletter
\newcommand*{\rom}[1]{\expandafter\@slowromancap\romannumeral #1@}
\makeatother
\newcommand{\angstrom}{\text{\normalfont\AA}}

\begin{document}

\preprint{APS/123-QED}

\title{Charge order induced Dirac pockets in the nonsymmorphic crystal TaTe$_4$}

\author{Yichen Zhang}
\thanks{These authors contributed equally to this work}
\affiliation{Department of Physics and Astronomy, Rice University, Houston, Texas 77005, USA}%

\author{Ruixiang Zhou}
\thanks{These authors contributed equally to this work}
\affiliation{School of Physical Science and Technology, Shanghai Tech University, Shanghai 200031, China}

\author{Hanlin Wu}
\thanks{These authors contributed equally to this work}
\affiliation{Department of Physics, The University of Texas at Dallas, Richardson, Texas 75080, USA}



\author{Ji Seop Oh}
\affiliation{Department of Physics and Astronomy, Rice University, Houston, Texas 77005, USA}
\affiliation{Department of Physics, University of California, Berkeley, California 94720, USA}

\author{Sheng Li}
\affiliation{Department of Physics, The University of Texas at Dallas, Richardson, Texas 75080, USA}

\author{Jianwei Huang}
\affiliation{Department of Physics and Astronomy, Rice University, Houston, Texas 77005, USA}

\author{Jonathan D. Denlinger}
\affiliation{Advanced Light Source, Lawrence Berkeley National Laboratory, Berkeley, California 94720, USA}

\author{Makoto Hashimoto}
\affiliation{Stanford Synchrotron Radiation Lightsource, SLAC National Accelerator Laboratory, Menlo Park, California 94025, USA}

\author{Donghui Lu}
\affiliation{Stanford Synchrotron Radiation Lightsource, SLAC National Accelerator Laboratory, Menlo Park, California 94025, USA}

\author{Sung-Kwan Mo}
\affiliation{Advanced Light Source, Lawrence Berkeley National Laboratory, Berkeley, California 94720, USA}

\author{Kevin F. Kelly}
\affiliation{Department of Electrical and Computer Engineering, Rice University, Houston, Texas, 77005, USA}

\author{Gregory T. McCandless}
\affiliation{Department of Chemistry and Biochemistry, Baylor University, Waco, Texas 76798, USA}

\author{Julia Y. Chan}
\affiliation{Department of Chemistry and Biochemistry, Baylor University, Waco, Texas 76798, USA}

\author{Robert J. Birgeneau}
\affiliation{Department of Physics, University of California, Berkeley, California 94720, USA}
\affiliation{Materials Science Division, Lawrence Berkeley National Laboratory, Berkeley, California 94720, USA}

\author{Bing Lv}
\email{blv@utdallas.edu}
\affiliation{Department of Physics, The University of Texas at Dallas, Richardson, Texas 75080, USA}

\author{Gang Li}
\email{ligang@shanghaitech.edu.cn}
\affiliation{School of Physical Science and Technology, Shanghai Tech University, Shanghai 200031, China}

\author{Ming Yi}
\email{mingyi@rice.edu}
\affiliation{Department of Physics and Astronomy, Rice University, Houston, Texas 77005, USA}



\date{\today}

\begin{abstract}
The interplay between charge order (CO) and nontrivial band topology has spurred tremendous interest in understanding topological excitations beyond the single-particle description. In a quasi-one-dimensional nonsymmorphic crystal TaTe$_4$, the (2a$\times$2b$\times$3c) charge ordered ground state drives the system into a space group where the symmetry indicators feature the emergence of Dirac fermions and unconventional double Dirac fermions. Using angle-resolved photoemission spectroscopy and first-principles calculations, we provide evidence of the CO induced Dirac fermion-related bands near the Fermi level. Furthermore, the band folding at the Fermi level is compatible with the new periodicity dictated by the CO, indicating that the electrons near the Fermi level follow the crystalline symmetries needed to host double Dirac fermions in this system. 
\end{abstract}

\maketitle
\section{\label{sec:level1}\textbf{\rom{1}. Introduction}\protect}

Since the discovery of Chern insulators~\cite{Haldane1988} and topological insulators~\cite{Kane2005,Kane2005_2}, fascinating physics about symmetry-protected topological phases has enormously expanded the research scope of quantum materials. The idea of defining bulk topological invariants was soon extended from gapped to gapless systems, featuring Dirac and Weyl semimetals~\cite{Armitage2018}. Furthermore, due to the nonsymmorphic crystalline symmetries in solid state systems, low energy electronic excitations in crystals can achieve, in addition to Dirac and Weyl fermions, unconventional three,four,six,and eightfold fermions~\cite{Wieder2016,Bradlyn2016,Tang2017} with no analog in the standard model. The experimental realization of ideal materials hosting these unconventional quasiparticles has been an ongoing task~\cite{Lv2017,Ma2018,Takane2019,Li2019,Sanchez2019,Schroter2019,Schroter2020,Lv2019,Yang2020,Kumar2020,Ju2022,Rong2022}. In particular, experimental efforts on realizing eightfold fermions robust against spin-orbit coupling (SOC) have been hindered by the restricted number of space groups~\cite{Wieder2016} and candidates strongly affected by electron-electron (e-e) interactions~\cite{Sante2017}. 

Beyond pristine lattices, charge orders (COs) and charge density wave (CDW) orders can further modify the crystalline symmetries. Originating from Peierl’s description of a one-dimensional (1D) chain~\cite{Peierls1955,Frohlich1954} where all electrons at the Fermi level (E$_{\rm F}$) can be connected with the same vector $q$=2$k_{\rm F}$ ($k_{\rm F}$ being the Fermi momentum), CDW order is the spontaneous translational symmetry breaking by the electronic degree of freedom, and can arise from Fermi surface nesting (FSN), electron-phonon coupling (EPC), and e-e interaction~\cite{Zhu2015}, where a weak instability tendency of FSN has been pointed out beyond one dimension~\cite{Johannes2008,Zhu2017}. CO shares the same phenomenology of modulated charge density upon a crystal lattice but differs in that the charge degree of freedom follows that of the lattice, which drives the translational symmetry breaking. More recently, CDWs and COs have received extensive attention due to their emergence in a series of correlated topological materials, including the possible axion physics in (TaSe$_4$)$_2$I~\cite{Gooth2019,Shi2021} and the COs found in the kagome metals, both in non-magnetic AV$_3$Sb$_5$ (A = K, Rb, Cs)~\cite{Li2020,Wu2022,Luo2022,Xie2022} and ScV$_6$Sn$_6$~\cite{Cheng2023}, as well as the magnetic kagome lattice FeGe~\cite{Teng2022,Teng2022_2,Yin2022}. Therefore, investigating the interplay between CO and nontrivial topological band structure is of particular interest. Here, we focus on the scenario where the CO modifies the crystalline symmetries to produce new topological crossings and potential unconventional eightfold fermions, as probed in the quasi-1D material TaTe$_4$.

The subcell of TaTe$_4$ crystalizes in the \textit{P}4/\textit{mcc} space group (No. 124), with Ta chains aligned along the [001] direction. TaTe$_4$ with its isostructural analogs NbTe$_4$~\cite{Bjerkelund1964,Boswell1983,Mahy1983,Selte1964} and Ta$_{1-x}$Nb$_x$Te$_4$~\cite{Eaglesham1985,Morelli1986,Boswell1986,Walker1988,Prodan1990,Bennett1992,Kusz1994,Kusz1995} form to a family of quasi-1D compounds that exhibit a series of incommensurate, commensurate, and discommensurate COs. In TaTe$_4$, a (2a$\times$2b$\times$3c) CO has been reported by single crystal x-ray and electron diffraction experiments~\cite{Bjerkelund1964,Boswell1983,Mahy1983,Bronsema1987,Bennett1991}, with an onset temperature of T$_{\rm CO}$ = 450 K. Within the CO state, transport studies show a metallic behavior~\cite{Tadaki1990}. Previous ARPES measurements, however, report a sizable gap at the Fermi level, E$_{\rm F}$, contradictory to the metallic transport behavior~\cite{Zwick1999}.  Magnetotransport measurements have revealed a large magnetoresistance and quantum oscillations~\cite{Tadaki1990,Sambongi1993,Luo2017,Gao2017,Zhang2020}. A nontrivial Berry phase was also suggested~\cite{Luo2017}, although more rigorous analysis involving higher-order harmonics~\cite{Alexandradinata2018} may be needed to pin down the Berry phase. Furthermore, recent scanning tunneling microscopy (STM) work revealed a (4a$\times$6c) surface CO~\cite{Sun2020} and edge states~\cite{Zhang2020}. The (4a$\times$6c) surface CO distinct from that of the bulk was attributed to the surface-enhanced e-e interaction, with EPC stabilizing CO states at high binding energies, while the 1D edge states were proposed to be evidence of the topological nature~\cite{Zhang2020} of TaTe$_4$. Regarding the bulk CO, recent first-principles calculations designated EPC the dominant role for triggering the phase transition, rather than FSN~\cite{Liu2021,Guster2022}.  Microscopically, the CO was proposed to be induced by the tendency to maximize Ta-Ta bonding, while introducing minimum shrinkage in the Te-sublattice bondings~\cite{Guster2022}. By using state-of-the-art ARPES, we first visualize the metallic band structure of TaTe$_4$, which resolves previous inconsistencies between ARPES spectra and optical measurements~\cite{Zwick1999}. In particular, our results provide spectroscopic evidence of the CO induced Dirac points (DPs) respecting the (2a$\times$3c) periodicity near E$_{\rm F}$. Such folded structure serves as the prerequisite for the nonsymmorphic symmetry-protected double Dirac fermions in the CO state of TaTe$_4$. Despite that the direct observation of the double Dirac points (DDPs) still awaits future efforts, our results provide a step towards identifying quantum materials where electronic orders can be utilized in introducing new symmetries to realize novel topological states.

\begin{figure*}[ht]
\centering
\includegraphics[width=1.0\textwidth]{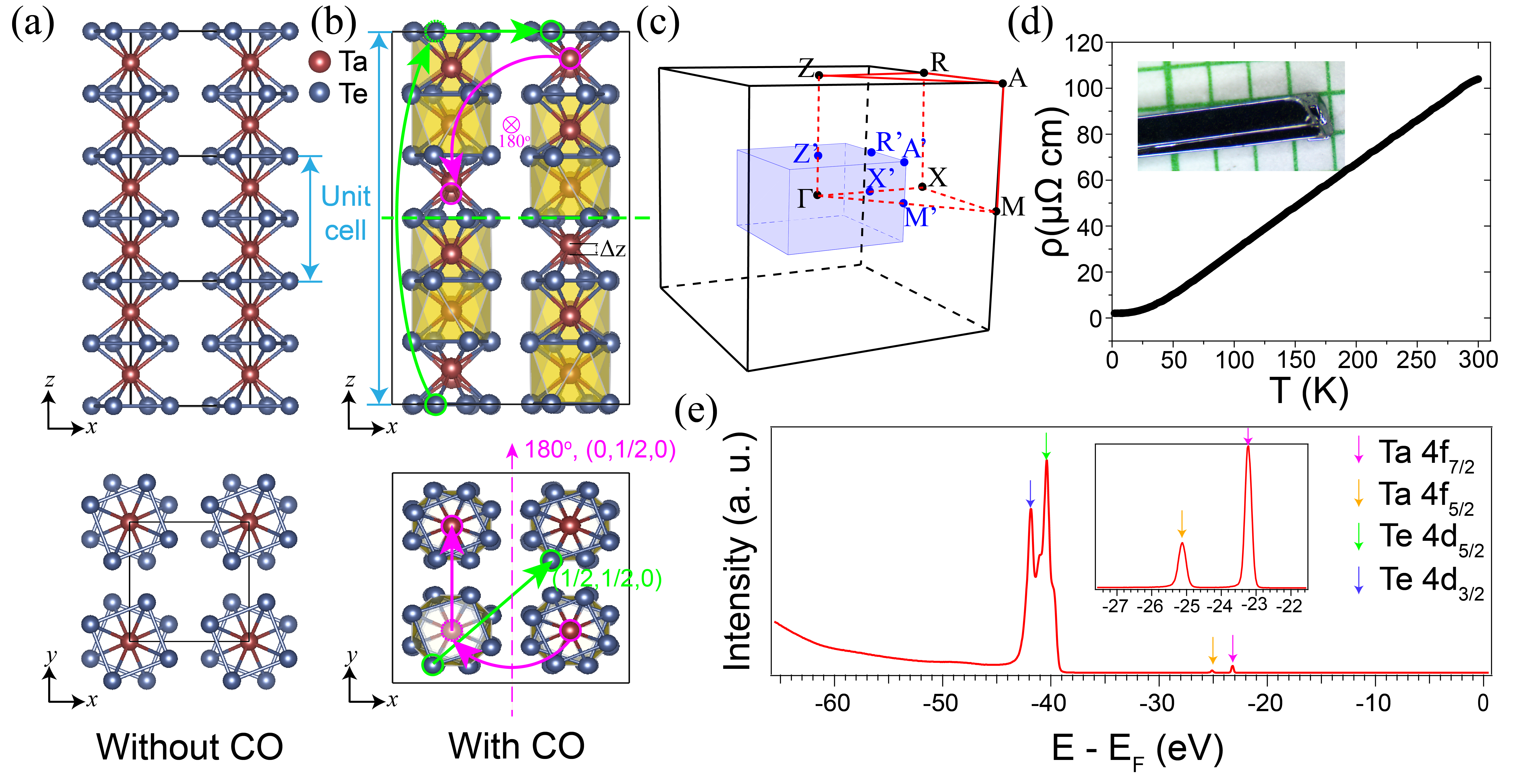}
\caption{Crystal structure, Brillouin zone, resistivity, and core level spectrum. (a) Side and top views of the subcell crystal structure of TaTe$_4$ in the non-CO phase (\textit{P}4/\textit{mcc}, No. 124). $x$, $y$, and $z$ correspond to the directions of the $a$, $b$, and $c$ lattice vectors. (b) Supercell crystal structure of TaTe$_4$ in the CO phase (\textit{P}4/\textit{ncc}, No. 130) with a twofold screw axis indicated by the magenta cross pointing into the page in the side view and the magenta arrows in the top view. A glide mirror plane is indicated by a horizontal green dashed line on the side view. Magenta and green circles illustrate the transformation of atom positions after the twofold screw axis and glide mirror symmetry operations, respectively. The dashed magenta and green circles serve as the intermediate positions before the fractional translations. (c) Non-CO BZ (black) and CO BZ (blue) of TaTe$_4$. The folded high symmetry points are denoted in blue and with a prime on the superscript. (d) Temperature dependent resistivity data. Inset is the image of TaTe$_4$ single crystal on a millimeter-scale sheet. (e) Core level spectrum of TaTe$_4$ highlighting Ta 4$f$ and Te 4$d$ orbitals in colored arrows.
}
\label{F1}
\end{figure*}

\section{\textbf{\rom{2}. Experimental methods}\protect}

\begin{figure*}[ht]
\centering
\includegraphics[width=1.0\textwidth]{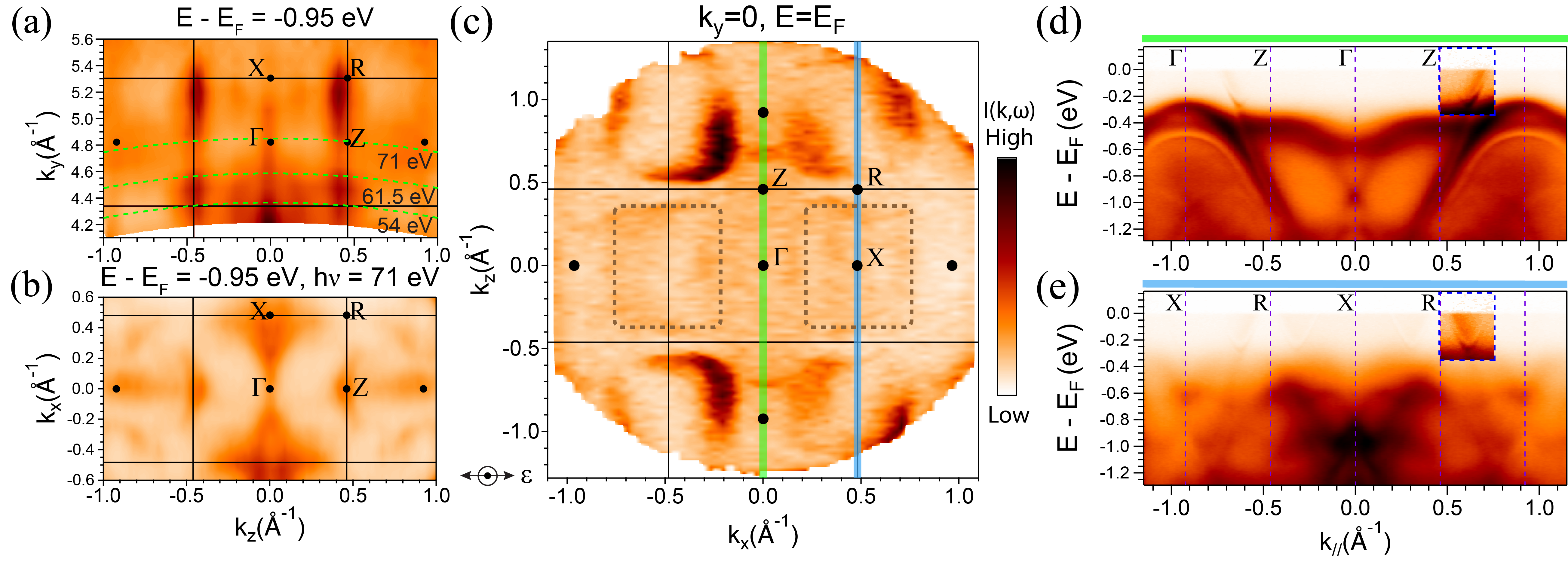}
\caption{Basic electronic band structure. (a) Photon-energy dependent constant energy contour at binding energy 0.95 eV where $k_y$ corresponds to the photon energy direction. 71 eV and 54 eV are found out to probe $k_y$ = 0 and $\pi$, respectively. (b) In-plane constant energy contour at the same binding energy as (a) to display similarities. (c) FS map using linear horizontal polarization at $k_y$ = 0 plane plotted in non-CO BZ. Dashed rectangles indicate pockets that are strongly suppressed by matrix element effect in the first BZ. (d), (e) Band dispersions along $\Gamma Z$ and $XR$, denoted by vertical green and blue bars in panel (c). Insets of (d) and (e) show the same region with enhanced color scale to emphasize the bands crossing the Fermi level. 
}
\label{F2}
\end{figure*}

Single crystals of TaTe$_4$ were grown using flux method. High-purity Ta powder (99.99\%, Alfa Asear) and Te lumps (99.9995\%, Alfa Asear) with molar ratio of 1:20 were placed into a quartz tube in a glove box with O$_2$ level and H$_2$O level $<$ 0.1 ppm. The quartz tube was flame sealed under vacuum. The quartz tube assembly was then heated in a box furnace from room temperature up to 1000 °C in 20 hours, stayed at 1000 °C for 24 hours and then slowly cooled to 550 °C at a rate of 3°C/h. Silver needle-like crystals with a typical size of 2$
\times$0.2$\times$0.2 mm$^3$ were obtained by decanting the flux with a centrifuge at 550 °C using quartz wool as a filter.

Single-crystal x-ray diffraction (XRD) was performed using a Bruker D8 Quest diffractometer equipped with a CMOS detector, and an Oxford Cryosystems 700 Series temperature controller. A hemisphere of frames was measured using a narrow-frame method with a scan width of 0.5° in $\varphi$ and $\omega$ and an exposure time of 12 s/frame with Mo K$\alpha$ radiation. The collected dataset was integrated using the Bruker Apex3 program, with the intensities corrected for the Lorentz factor, polarization, air absorption, and absorption due to variation in the path length through the detector faceplate. The data were scaled, and absorption correction was applied using SADABS~\cite{Krause2015}. A starting model was obtained using the intrinsic method in SHELXT~\cite{Sheldrick2015}, and atomic sites were refined anisotropically using SHELXL2014/7. Powder XRD was carried out at room temperature on a Rigaku Smartlab x-ray diffractometer with a monochromatic Cu K$\alpha$1 radiation. Electrical resistivity $\rho$(T,H) using a four probe configuration was measured down to 2 K and up to 9 T magnetic field in a quantum design physical property measurement system (PPMS). The composition and homogeneity of the single crystals were confirmed by scanning electron microscopy with energy dispersive x-ray spectroscopy (SEM-EDX) and mapping using Zeiss EVO LS 15 SEM with 20 kV.

ARPES measurements were performed at the Advanced Light Source, Beamlines 4.0.3 and 10.0.1, equipped with SCIENTA R8000 and R4000 analyzers, respectively, and at the Stanford Synchrotron Radiation Lightsource, BL 5-2, equipped with a DA30 electron analyzer. The single crystals were cleaved \textit{in-situ} at 10 K and measured in ultra-high vacuum with a base pressure better than 5$\times$10$^{-11}$ Torr. Photon energy-dependent measurements cover from 30 to 200 eV.  Energy and angular resolution were set to be better than 25 meV and 0.1°, respectively.

Charge self-consistent DFT calculations were performed with the experimental lattice constants by using the generalized gradient approximation (GGA) potential~\cite{PBE1996} as implemented in the Vienna \textit{ab initio} simulation package~\cite{Kresse1996,Kresse1996_2}. A $k$ mesh of 7$
\times$7$\times$7 for the non-CO phase and 2$\times$2$\times$1 for the CO phase was adopted. To better account for the interaction effect poorly approximated in GGA, we further employed the HSE06 functional to calculate the electronic structure along $\Gamma Z$ and $XR$ (see Fig.~S5 in the Supplemental Material~\cite{SM}). The CO electronic structure folds back to the BZ of the non-CO state via the vaspkit code~\cite{Wang2021}.

\section{\textbf{\rom{3}. Results and discussion}\protect}

\begin{figure*}[ht]
\centering
\includegraphics[width=0.7\textwidth]{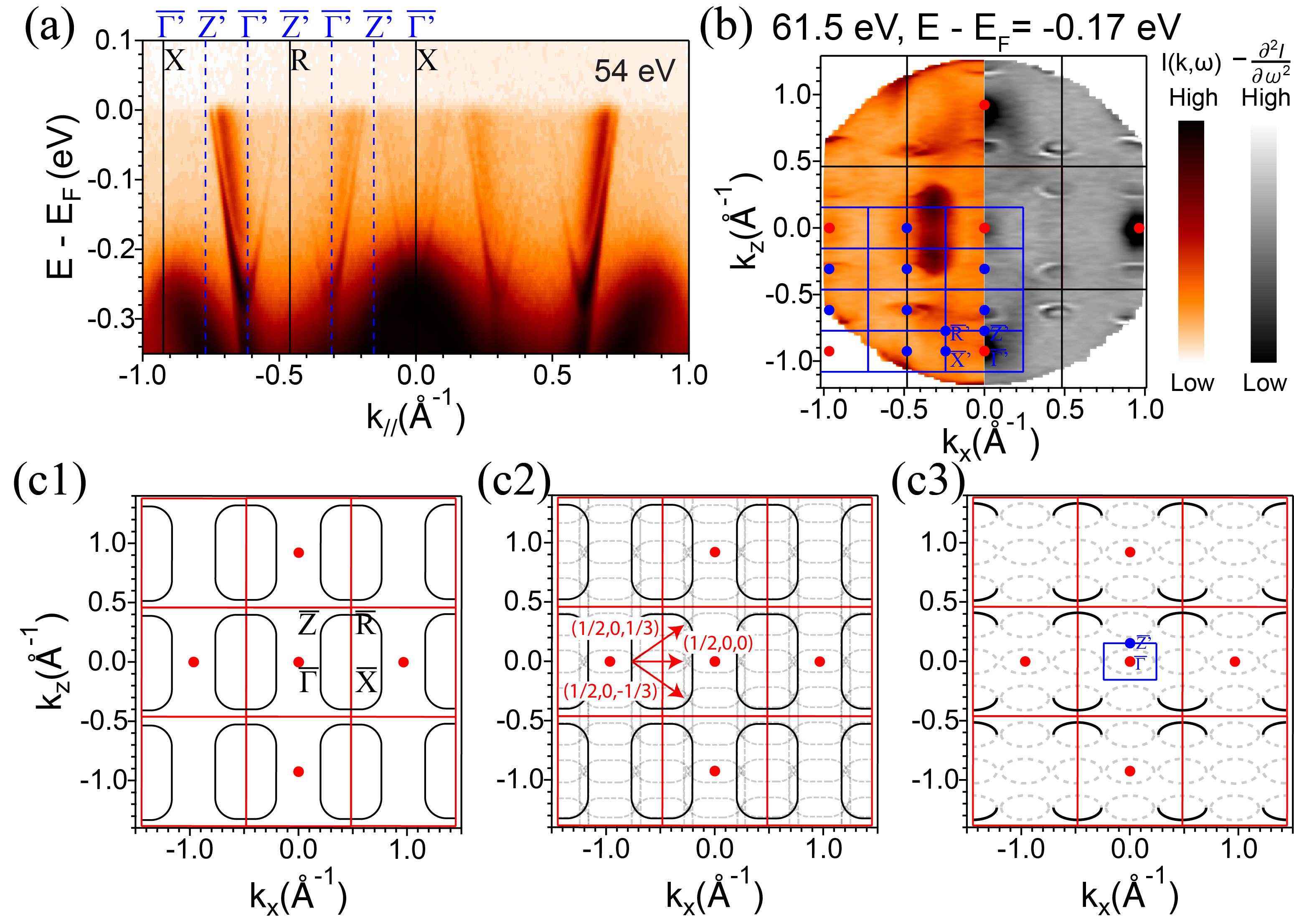}
\caption{CO folded bands near E$_{\rm F}$ and folding mechanism. (a) Zoom-in on band dispersions along $XR$ ($\Gamma'Z'$) near E$_{\rm F}$ showing the sharp outer and the broadened inner electron pockets obeying CO BZ periodicity. (b) Constant energy contour of the $\overline{\Gamma'}$-$\overline{Z'}$-$\overline{R'}$-$\overline{X'}$ plane at E - E$_{\rm F}$ = -0.17 eV. Raw data (from right half) mirrored to the left half and second derivative along energy for better visualization of the CO-folded pockets on the right half. (c1)-(c3) Schematic illustrations to qualitatively explain the folding mechanism of the Fermi pockets. Solid and dashed transparent rectangles in (c1) and (c2) are original and folded pockets translated by $\bm{q}_{\rm CO}$, respectively. Final folded ellipses in non-CO and CO Brillouin zones are shown in (c3).   
}
\label{F3}
\end{figure*}

As the crystalline symmetries are important for the topological properties, we begin by discussing the crystal structure of the pristine and CO phases. The subcell crystal structure of non-CO TaTe$_4$ consists of tantalum chains surrounded by tellurium antiprisms [Fig.~\ref{F1}(a)]. In the CO phase (space group \textit{P}4/\textit{ncc}, No. 130), the distorted unit cell is tripled along $z$ and doubled along the two equivalent $x$ and $y$ directions [Fig.~\ref{F1}(b)]. The lattice parameters at T = 100 K in the CO phase identified from single crystal x-ray diffraction are: $a$ = $b$ = 12.996(3) $\angstrom$ and $c$ = 20.385(5) $\angstrom$ (See Table.~S1 in the Supplemental Material~\cite{SM}). The CO introduces new nonsymmorphic symmetries: for example, a twofold screw symmetry along $y$ and a glide mirror symmetry about the $xy$ plane. The two symmetry operations are illustrated by the magenta and green symbols in Fig.~\ref{F1}(b), respectively. When transforming to reciprocal space, the CO shrinks the pristine state BZ by a factor of (2$\times$2$\times$3), as depicted in Fig.~\ref{F1}(c). Throughout the paper, we label high symmetry points in the CO BZ in primed notation to distinguish them from their non-CO counterparts. TaTe$_4$ exhibits good metallic behavior, as indicated by the temperature-dependent resistivity with a relatively high residue resistivity ratio (RRR = 51) [Fig.~\ref{F1}(d)], indicating the good quality of grown single crystals. Further characterization of the single crystal magnetoresistance and homogeneity is presented in Fig.~S1 and Fig.~S2. In addition, core-level x-ray photoelectron spectrum identifies both Ta 4$f$ and Te 4$d$ peaks [Fig.~\ref{F1}(e)]. Since the natural cleaving plane for TaTe$_4$ is the (010) plane, $xz$ ($ac$) is the experimental in-plane direction, while the $y$ direction is the out-of-plane direction in the ARPES experiments. We also note that due to the equivalence of the $a$ and $b$ axes of the crystal, $k_x$ and $k_y$ are equivalent for the bulk crystal. 

\begin{figure*}[ht]
\centering
\includegraphics[width=0.9\textwidth]{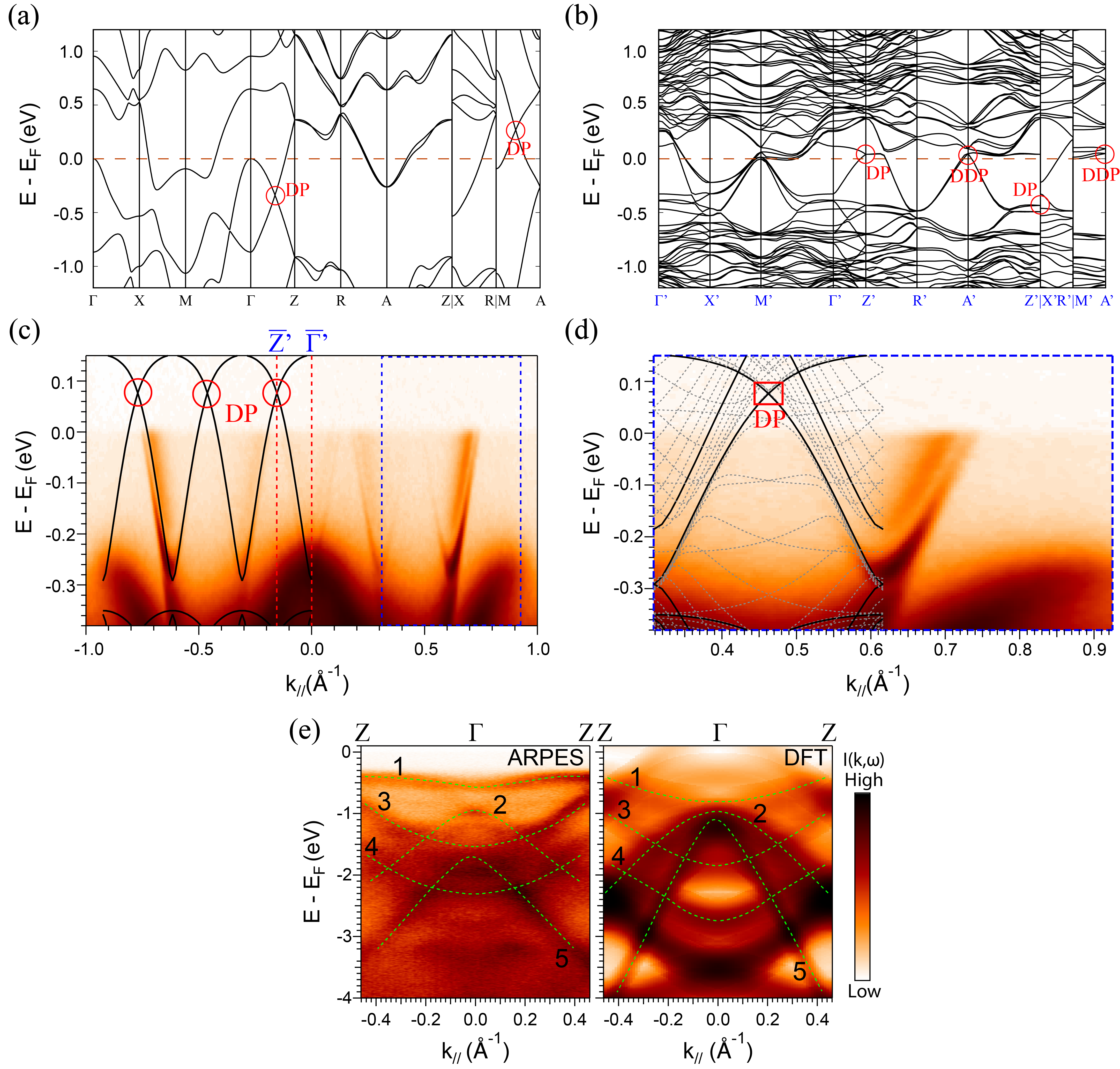}
\caption{CO induced DPs and ARPES spectra in comparison with DFT calculations. (a) First-principles calculations of the electronic band structure in the non-CO state with SOC. (b) Same calculation considering SOC but in the CO ground state. Both (a) and (b) have no chemical potential shift. (c) $\overline{\Gamma'}$-$\overline{Z'}$ band dispersions (54 eV) with bulk first-principles calculations in the CO phase (black lines) overlaid on the experimental data. E$_{\rm F}$ of the calculation is shifted down by 35 meV to achieve the best match with ARPES data. Red circles highlight the CO induced DPs above E$_{\rm F}$. (d) Zoom-in Dirac band structure from the region boxed by blue dashed lines in (c). A $k_y$ broadened calculation with 35 meV downshift of E$_{\rm F}$ is superimposed, where black solid lines are bands calculated from $\Gamma'Z'$ ($k_y$=0) and $X'R'$ [$k_y$=0.5$\times$(2$\pi/b$)], while gray dotted bands are from nine equally spaced $k_y$ values between 0 and 0.5$\times$(2$\pi/b$). $b$ is a lattice vector of the CO supercell. (e) Large energy range comparison between backfolded first-principles calculation (right) and ARPES measured dispersions (left) along $\Gamma Z$ (non-CO BZ). Green dashed lines are to show qualitative similarities of bands one to five identified in ARPES and DFT. No chemical potential shift is applied in (e). DP: Dirac point. DDP: Double Dirac point.
}
\label{F4}
\end{figure*}

We first present the overall electronic structure measured by ARPES. To determine the inner potential, we compare the $k_y$-$k_z$ map measured in the out-of-plane direction by varying photon energy [Fig.~\ref{F2}(a)] and the equivalent $k_x$-$k_z$ map measured in the in-plane direction by varying angle at the same energy (E - E$_{\rm F}$ = -0.95 eV). The bulk states should be equivalent in these two maps, due to the equivalence of $k_x$ and $k_y$. From this comparison, we can identify the periodicity along $k_y$ from a dispersive feature along $\Gamma X$ where the narrowest (widest) part is located at $\Gamma$ ($X$), giving us an inner potential V$_0$ of 23 eV. From the Fermi surface (FS) measured at $k_y$ = 0 [Fig.~\ref{F2}(c)], we observe rectangular pockets centered around the $X$ points, with a periodicity compatible with the non-CO BZ, albeit strongly suppressed in the first BZ due to photoemission matrix element effects. Hence, these features must be associated with the band structure of the non-CO state. We also map out dispersions along the $\Gamma$-$Z$ and $X$-$R$ directions in Figs.~\ref{F2}(d) and (e). Spectral intensity is observed mostly beyond 0.2 eV of E$_{\rm F}$, consistent with earlier identification of the CO gap in previous ARPES results~\cite{Zwick1999}. However, in contrast to earlier reports where no dispersions were observed within the 0.2 eV gap, we clearly resolve the presence of dispersions up to E$_{\rm F}$, reconciling the inconsistency with the metallic transport and optical conductivity behavior. From a more detailed view near E$_{\rm F}$ with an enhanced color scale, the near-E$_{\rm F}$ feature consists of a sharp outer electron pocket, with broadened intensity inside. In the following, we focus on these fine features near E$_{\rm F}$.

As shown in Fig.~\ref{F3}(a), the near-E$_{\rm F}$ electronic band structure probed by 54 eV photons shows multiple electronlike pockets which follow the folded BZ periodicity along $\overline{\Gamma'}\overline{Z'}$. They consist of two characteristic features: a sharp outer electron pocket and broadened inner electron pockets. To better illustrate the (2a$\times$3c) CO folding, we performed second derivative analysis (along the energy direction) on the constant energy contour at E - E$_{\rm F}$ = -0.17 eV, $h\nu$ = 61.5 eV, which nominally corresponds to the $k_y$ = $\pi$ plane ($X'R'A'M'$ plane) for the CO BZ. However, we note here that due to the low $k_y$ (out-of-plane) resolution (on the order of 0.1 $\angstrom^{-1}$)~\cite{Sobota2021} in the vacuum ultraviolet (VUV) regime of ARPES, it is challenging to precisely distinguish the $k_y$ = 0 or $\pi$ planes separated by 0.15 $\angstrom^{-1}$ in the CO BZ. Therefore, we label the high symmetry plane as $\overline{\Gamma'}\overline{Z'}\overline{R'}\overline{X'}$ in Fig.~\ref{F3}(b). On the left half of Fig.~\ref{F3}(b), the mirrored raw intensity of the map shows both the rectangular Fermi pockets from the non-CO BZ similar to the ones marked in Fig.~\ref{F2}(c), and the small elliptical pockets near the boundaries of the rectangular pockets. These small elliptical pockets can be seen more clearly in the second derivative plot on the right half of the map, where they are centered at the $\overline{\Gamma'}$ points of the CO folded BZ. Examining the ellipses in greater detail, one can see the small pockets to consist of two parts: the sharp outer components (white ellipses) and the broadened inner part (inner white intensities elongated along horizontal direction) corresponding to the features discussed in Fig.~\ref{F3}(a). Additionally, we do not observe any evidence of the (4a$\times$6c) surface CO suggested in a recent STM work~\cite{Zhang2020,Sun2020}. 

Next we illustrate the connection between the unfolded rectangular pockets and the folded elliptical pockets in the CO BZ [Figs.~\ref{F3}(c1)-(c3)]. If we start with the unfolded rectangular pockets as observed in the non-CO BZ [red BZ grids in Figs.~\ref{F3}(c1)-(c3)], we note that the folding of the pockets across $\bm{q}_{\rm CO}=(\pm\frac{1}{2}, 0, \pm\frac{1}{3})$ and its subset would indeed lead to the overlap of the rectangular pockets, as indicated by the interconnected dashed rectangles in Fig.~\ref{F3}(c2), the hybridization of which would produce elliptical pockets at the center of the CO-folded BZ, as shown in Fig.~\ref{F3}(c3). Here, we denote the ellipses from folding in dashed style, while the ones shared with original rectangular pockets in solid ellipses.
Such band folding scheme provides a qualitative understanding of the relation between the observed unfolded rectangular pockets and the folded elliptical pockets, and confirms that the near-E$_{\rm F}$ electronic structure obeys the periodicity of the CO folding. 

Since the CO introduces new nonsymmorphic symmetries to the crystal structure, we expect new topological features to arise due to the CO. In order to gain insights into the electronic structure from the theory perspective, we carried out DFT calculations with different exchange-correlation functionals. In Fig.~\ref{F4}, we present the main results from the generalized gradient approximation (GGA) by Perdew, Burke, and Ernzerhof (PBE) parameterization, while attempts on the HSE06 functional in comparison with ARPES data can be found in Fig.~S5~\cite{SM}. Regarding the PBE calculation, the electronic structure of both the non-CO (Fig.~\ref{F4}(a)) and CO [Fig.~\ref{F4}(b)] states shows clear metallic nature with both electron and hole pockets. The charge modulation in TaTe$_4$ does not lead to gap openings at the Fermi level, consistent with the experimental observations. As indicated by other calculations~\cite{Zhang2020,Liu2021,Guster2022}, linear band crossings exist along $\Gamma Z$ and $MA$ in the non-CO phase, as circled in Fig.~\ref{F4}(a), leading to DPs. The presence of the CO modifies the band topology. In addition to DPs, the CO introduces double Dirac points (DDPs) pinned at the $A'$ point~\cite{Wieder2016} [Fig.~\ref{F4}(b)]. The symmetry protection mechanisms of DPs in the two phases of TaTe$_4$ are slightly different. In the non-CO phase, DPs appear along high-symmetry lines, while in the CO phase, DPs are pinned at the time-reversal invariant momentum (TRIM) points.

Finally, we report on the direct observation of the CO induced Dirac point (DP)-related bands near E$_{\rm F}$, assisted by the comparison with DFT calculations. As shown in Fig.~\ref{F4}(c), the $\overline{\Gamma'}$-$\overline{Z'}$ dispersions of the near-E$_{\rm F}$ electron bands agree reasonably well with the bulk DFT calculations for the CO state after a 35 meV downshift of E$_{\rm F}$. In particular, the periodicity of this band is consistent with the CO phase. Interestingly, DFT calculation shows these bands to cross in the form of DPs at about 75 meV above the Fermi level. Given that these are the dominant dispersions we observe near E$_{\rm F}$, these DPs may potentially be the source of the nontrivial Berry phase, large magnetoresistance and resistivity plateau observed in recent transport measurements~\cite{Luo2017,Gao2017}. To further analyze the details of the DP-related bands, we zoom in to the blue boxed region of Fig.~\ref{F3}(c) and overlay the $k_y$-broadened bulk DFT calculations between $\Gamma' Z'$ and $X'R'$ (see Fig.~S4~\cite{SM} for detailed contributions from different $k_y$ planes) on top of the ARPES spectra in Fig.~\ref{F4}(d), where the two separated features of the broadened inner and sharp outer pockets can be well explained by an out-of-plane momentum collapse of the photoemission data, reflecting the $k_y$ distribution of the DFT calculated bands. Considering the aforementioned out-of-plane momentum resolution of VUV ARPES, the explanation is further justified. However, the predicted DDPs at deeper binding energies are not directly observed (see Fig.~S6~\cite{SM}). The underlying reason might be a combination of the following three factors. First, VUV ARPES may not have sufficient out-of-plane resolution to selectively probe the DDPs at $A'$. Second, the CO folded bands, band hybridization, and band backfolding onto the non-CO BZ could cause further broadening effect on ARPES spectra, which hinders the task of resolving a high-fold band degeneracy up to eight. Third, we note that away from the near-E$_{\rm F}$ features, neither the non-CO nor the CO state DFT calculations show good agreement with ARPES spectra, as detailed in Fig.~S6. Such mismatch impedes a directly experimental identification of the DDPs at deeper binding energies. The reason for such inconsistency is subject to future studies, while the recent STM work~\cite{Sun2020} has suggested complicated surface reconstruction and possible enhanced surface e-e interactions, which are not captured by our current DFT calculations. Nonetheless, we find that by backfolding the CO state calculations to non-CO BZ, an overall better agreement between DFT and ARPES can be obtained, as exemplified in Fig.~\ref{F4}(e) for $\Gamma Z$. Specifically, one can observe a qualitative match for bands one to five as denoted, but unlike a good theoretical description for the near-E$_{\rm F}$ electron pockets, the positions and slopes of these bands at deeper binding energies do not coincide precisely.

\section{\textbf{\rom{4}. Conclusion}\protect}

In summary, we have presented ARPES measurements of the electronic structure on the CO compound TaTe$_4$. We clearly reveal the presence of low energy dispersions that reconcile the inconsistency between previous ARPES report and transport results. In particular, we directly visualize the CO-folded electron pockets near E$_{\rm F}$, which are interpreted as the consequence of CO folding via the known CO wavevector $\bm{q}_{\rm CO}$ = ($\frac{1}{2}$, 0, $\frac{1}{3}$) from the original bands. Moreover, we present DFT calculations to understand the CO band structure and the role of topology in the nonsymmorphic crystal, where the DPs and DDPs are induced due to the emergent crystalline symmetries of the CO. The PBE calculations in the CO state with a $k_y$ stacking show good agreements for the near-E$_F$ $\overline{\Gamma'}$-$\overline{Z'}$ electron pockets, indicating the direct observation of the lower branch of the symmetry-enforced DPs. Despite that the predicted DDPs remain elusive in terms of experimental visualization, TaTe$_4$ presents an interesting material platform where CO induces new crystalline symmetries that result in Dirac quasiparticles near the Fermi level.


\section{acknowledgments}
This research used resources of the Advanced Light Source and the Stanford Synchrotron Radiation Lightsource, both U.S. DOE Office of Science User Facilities under Contract No. DE-AC02-05CH11231 and No. AC02-76SF00515, respectively. ARPES work at Rice University is supported by the Department of Defense, Air Force Office of Scientific Research under Grant No. FA9550-21-1-0343; and the Robert A. Welch Foundation grant no. C-2175. J.S.O. is supported by NSF Grant No. DMR-1921847 and No. DMR-1921798.
Work at the University of California, Berkeley, is funded by the U.S. Department of Energy, Office of Science, Office of Basic Energy Sciences, Materials Sciences and Engineering Division under Contract No. DE-AC02-05-CH11231 (Quantum Materials program KC2202). Synthesis work at UT Dallas is supported by NSF under Grant No.
DMR-1921581, AFOSR under Grant No. FA9550-19-1-0037
and ONR under Grant No. N00014-23-1-2020. X-ray structure determination and analysis is supported by Welch Foundation Welch AA-2056-20220101. G.L. acknowledges the National Key R\&D Program of China (2022YFA1402703), National Natural Science Foundation of China (11874263), Shanghai 2021-Fundamental Research Aera (21JC1404700), Shanghai Technology Innovation Action Plan (20DZ1100605), and Sino-German Mobility program (M-0006). M.H. and D.L. acknowledge the support of the U.S. Department of Energy, Office of Science, Office of Basic Energy Sciences, Division of Material Sciences and Engineering, under Contract No. DE-AC02-76SF00515.


\begin{thebibliography}{100}

\bibitem{Haldane1988} F. D. M. Haldane, \href{https://journals.aps.org/prl/abstract/10.1103/PhysRevLett.61.2015}{Phys. Rev. Lett. \textbf{61}, 2015-2018 (1988).} 

\bibitem{Kane2005} C. L. Kane and E. J. Mele, \href{https://journals.aps.org/prl/abstract/10.1103/PhysRevLett.95.226801}{Phys. Rev. Lett. \textbf{95}, 226801 (2005).} 

\bibitem{Kane2005_2} C. L. Kane and E. J. Mele, \href{https://journals.aps.org/prl/abstract/10.1103/PhysRevLett.95.146802}{Phys. Rev. Lett. \textbf{95}, 146802 (2005).} 

\bibitem{Armitage2018} N. P. Armitage, E. J. Mele and A. Vishwanath, \href{https://journals.aps.org/rmp/abstract/10.1103/RevModPhys.90.015001}{Rev. Mod. Phys. \textbf{90}, 015001 (2018).} 

\bibitem{Wieder2016} B. J. Wieder, Y. Kim, A. M. Rappe and C. L. Kane, \href{https://journals.aps.org/prl/abstract/10.1103/PhysRevLett.116.186402}{Phys. Rev. Lett. \textbf{116}, 186402 (2016).} 

\bibitem{Bradlyn2016} B. Bradlyn, J. Cano, Z. Wang, M. G. Vergniory, C. Felser, R. J. Cava and B. A. Bernevig, \href{https://www.science.org/doi/10.1126/science.aaf5037}{Science \textbf{353}, aaf5037 (2016).} 

\bibitem{Tang2017} P. Tang, Q. Zhou and S. C. Zhang, \href{https://journals.aps.org/prl/abstract/10.1103/PhysRevLett.119.206402}{Phys. Rev. Lett. \textbf{119}, 206402 (2017).}

\bibitem{Lv2017} B. Q. Lv, Z.-L. Feng, Q.-N. Xu, X. Gao, J.-Z Ma, L.-Y. Kong, P. Richard, Y.-B. Huang, V. N. Strocov, C. Fang, H.-M. Weng, Y.-G. Shi, T. Qian and H. Ding, \href{https://www.nature.com/articles/nature22390}{Nature \textbf{546}, 627-631 (2017).}

\bibitem{Ma2018} J.-Z. Ma, J.-B. He, Y.-F. Xu, B. Q. Lv, D. Chen, W.-L. Zhu, S. Zhang, L.-Y. Kong, X. Gao, L.-Y. Rong, Y.-B. Huang, P. Richard, C.-Y. Xi, E. S. Choi, Y. Shao, Y.-L. Wang, H.-J. Gao, X. Dai, C. Fang, H.-M. Weng, G.-F. Chen, T. Qian and H. Ding, \href{https://www.nature.com/articles/s41567-017-0021-8}{Nat. Phys. \textbf{14}, 349-354 (2018).}

\bibitem{Takane2019} D. Takane, Z. Wang, S. Souma, K. Nakayama, T. Nakamura, H. Oinuma, Y. Nakata, H. Iwasawa, C. Cacho, T. Kim, K. Horiba, H. Kumigashira, T. Takahashi, Y. Ando and T. Sato, \href{https://journals.aps.org/prl/abstract/10.1103/PhysRevLett.122.076402}{Phys. Rev. Lett. \textbf{122}, 076402 (2019).}

\bibitem{Li2019} H. Li, S. Xu, Z.-C. Rao, L.-Q. Zhou, Z.-J. Wang, S.-M. Zhou, S.-J. Tian, S.-Y. Gao, J.-J. Li, Y.-B. Huang, H.-C. Lei, H.-M. Weng, Y.-J. Sun, T.-L. Xia, T. Qian and H. Ding, \href{https://www.nature.com/articles/s41467-019-13435-4}{Nat. Commun. \textbf{10}, 5505 (2019).}

\bibitem{Sanchez2019} D. S. Sanchez, I. Belopolski, T. A. Cochran, X. Xu, J.-X. Yin, G. Chang, W. Xie, K. Manna, V. S\"{u}ß, C.-Y. Huang, N. Alidoust, D. Multer, S. S. Zhang, N. Shumiya, X. Wang, G.-Q. Wang, T.-R. Chang, C. Felser, S.-Y. Xu, S. Jia, H. Lin and M. Z. Hasan, \href{https://www.nature.com/articles/s41586-019-1037-2}{Nature \textbf{567}, 500-505 (2019).}

\bibitem{Schroter2019} N. B. M. Schr\"{o}ter, D.Pei, M. G. Vergniory, Y. Sun, K. Manna, F. de Juan, J. A. Krieger, V. S\"{u}ss, M. Schmidt, P. Dudin, B. Bradlyn, T. K. Kim, T. Schmitt, C. Cacho, C. Felser, V. N. Strocov and Y. Chen, \href{https://www.nature.com/articles/s41567-019-0511-y}{Nat. Phys. \textbf{15}, 759-765 (2019).}

\bibitem{Schroter2020} N. B. M. Schr\"{o}ter, S. Stolz, K. Manna, F. de Juan, M. G. Vergniory,J. A. Krieger, D. Pei, T. Schmitt, P. Dudin, T. K. Kim, C. Cacho,B. Bradlyn, H. Borrmann, M. Schmidt, R. Widmer, V. N. Strocov and C. Felser, \href{https://www.science.org/doi/10.1126/science.aaz3480}{Science \textbf{369}, 179-183 (2020).}

\bibitem{Lv2019} B. Q. Lv, Z.-L. Feng, J.-Z. Zhao, Noah F. Q. Yuan, A. Zong, K. F. Luo, R. Yu, Y.-B. Huang, V. N. Strocov, A. Chikina, A. A. Soluyanov, N. Gedik, Y.-G. Shi, T. Qian and H. Ding, \href{https://journals.aps.org/prb/abstract/10.1103/PhysRevB.99.241104}{Phys. Rev. B \textbf{99}, 241104(R) (2019).}

\bibitem{Yang2020} X. Yang, T. A. Cochran, R. Chapai, D. Tristant, J.-X. Yin, I. Belopolski, Z. Cheng, D. Multer, S. S. Zhang, N. Shumiya, M.Litskevich, Y. Jiang, G. Chang, Q. Zhang, I. Vekhter, W. A. Shelton, R. Jin, S.-Y. Xu and M. Z. Hasan, \href{https://journals.aps.org/prb/abstract/10.1103/PhysRevB.101.201105}{Phys. Rev. B \textbf{101}, 201105 (2020).}

\bibitem{Kumar2020} N. Kumar, M. Yao, J. Nayak, M. G. Vergniory, J. Bannies, Z. Wang, N. B. M. Schröter, V. N. Strocov, L. M\"{u}chler, W. Shi, E. D. L. Rienks, J. L. Mañes, C. Shekhar, S. S. P. Parkin, J. Fink, G. H. Fecher, Y. Sun, B. A. Bernevig and C. Felser, \href{https://onlinelibrary.wiley.com/doi/full/10.1002/adma.201906046}{Adv. Mater. \textbf{32}, 1906046 (2020).}

\bibitem{Ju2022} W. Ju, J.Jeong, E.-J. Cho, H.-J. Noh, K. Kim and B.-G. Park, \href{https://journals.aps.org/prb/abstract/10.1103/PhysRevB.106.205125}{Phys. Rev. B. \textbf{106}, 205125 (2022).}

\bibitem{Rong2022} H. Rong, Z. Huang, X. Zhang, S. Kumar, F.Zhang, C. Zhang, Y. Wang, Z. Hao, Y. Cai, L. Wang, C. Liu, X.-M. Ma, S. Guo, B. Shen, Y. Liu, S. Cui, K. Shimada, Q. Wu, J. Lin, Y. Yao, Z. Wang, H. Xu and C. Chen, \href{https://www.nature.com/articles/s41535-023-00565-8}{npj Quantum Mater. \textbf{8}, 29 (2023).}

\bibitem{Sante2017} D. D. Sante, A. Hausoel, P. Barone, J. M. Tomczak, G. Sangiovanni, and R. Thomale, \href{https://journals.aps.org/prb/abstract/10.1103/PhysRevB.96.121106}{Phys. Rev. B. \textbf{96}, 121106(R) (2017).}

\bibitem{Peierls1955} R. E. Peierls, \textit{Quantum theory of solids} (Oxford Univ. Press, Oxford, 1955).

\bibitem{Frohlich1954} H. Fr\"{o}hlich, \href{https://royalsocietypublishing.org/doi/10.1098/rspa.1954.0116}{Proc. R. Soc. Lond. A. \textbf{223}, 296-305 (1954).}

\bibitem{Zhu2015} X. Zhu, Y. Cao, J. Zhang, E. W. Plummer, and J. Guo, \href{https://www.pnas.org/doi/10.1073/pnas.1424791112}{Proc. Natl. Acad. Sci. U. S. A. \textbf{112}, 2367-2371 (2015).}

\bibitem{Johannes2008} M. D. Johannes and I. I. Mazin, \href{https://journals.aps.org/prb/abstract/10.1103/PhysRevB.77.165135}{Phys. Rev. B \textbf{77}, 165135 (2008).}

\bibitem{Zhu2017} X. Zhu, J. Guo, J. Zhang, and E. W. Plummer, \href{https://www.tandfonline.com/doi/full/10.1080/23746149.2017.1343098}{Adv. Phys. X. \textbf{2}, 622-640 (2017).}

\bibitem{Gooth2019} J. Gooth, B. Bradlyn, S. Honnali, C. Schindler, N. Kumar, J. Noky, Y. Qi, C. Shekhar, Y. Sun, Z. Wang, B. A. Bernevig and C. Felser, \href{https://www.nature.com/articles/s41586-019-1630-4}{Nature \textbf{575}, 315-319 (2019).}

\bibitem{Shi2021} W. Shi, B. J. Wieder, H. L. Meyerheim, Y. Sun, Y. Zhang, Y. Li, L. Shen, Y. Qi, L. Yang, J. Jena, P. Werner, K. Koepernik, S. Parkin, Y. Chen, C. Felser, B. A. Bernevig and Z. Wang, \href{https://www.nature.com/articles/s41567-020-01104-z}{Nat. Phys. \textbf{17}, 381-387 (2021).}

\bibitem{Li2020} H. Li, T. T. Zhang, T. Yilmaz, Y. Y. Pai, C. E. Marvinney, A. Said, Q. W. Yin, C. S. Gong, Z. J. Tu, E. Vescovo, C. S. Nelson, R. G. Moore, S. Murakami, H. C. Lei, H. N. Lee, B. J. Lawrie, and H. Miao \href{https://journals.aps.org/prx/abstract/10.1103/PhysRevX.11.031050}{Phys. Rev. X. \textbf{11}, 031050 (2021).}

\bibitem{Wu2022} S. Wu, B. R. Ortiz, H. Tan, S. D. Wilson, B. Yan, T. Birol, and G. Blumberg, \href{https://journals.aps.org/prb/abstract/10.1103/PhysRevB.105.155106}{Phys. Rev. B. \textbf{105}, 155106 (2022).}

\bibitem{Luo2022} H. Luo, Q. Gao, H. Liu, Y. Gu, D. Wu, C. Yi, J. Jia, S. Wu, X. Luo, Y. Xu, L. Zhao, Q. Wang, H. Mao, G. Liu, Z. Zhu, Y. Shi, K. Jiang, J. Hu, Z. Xu, and X. J. Zhou, \href{https://www.nature.com/articles/s41467-021-27946-6}{Nat. Commun. \textbf{13}, 273 (2022).}

\bibitem{Xie2022} Y. Xie, Y. Li, P. Bourges, A. Ivanov, Z. Ye, J.-X. Yin, M. Z. Hasan, A. Luo, Y. Yao, Z. Wang, G. Xu, and P. Dai, \href{https://journals.aps.org/prb/abstract/10.1103/PhysRevB.105.L140501}{Phys. Rev. B. \textbf{105}, L140501 (2022).}

\bibitem{Cheng2023} S. Cheng, Z. Ren, H. Li, J. Oh, H. Tan, G. Pokharel, J. M. DeStefano, E. Rosenberg, Y. Guo, Y. Zhang, Z. Yue, Y. Lee, S. Gorovikov, M. Zonno, M. Hashimoto, D. Lu, L. Ke, F.Mazzola, J. Kono, R. J. Birgeneau, J.-H. Chu, S. D. Wilson, Z. Wang, B. Yan, M. Yi, I. Zeljkovic, \href{https://arxiv.org/abs/2302.12227}{arXiv. 2302. 12227 (2023).}

\bibitem{Teng2022} X. Teng, L. Chen, F. Ye, E. Rosenberg, Z. Liu, J.-X. Yin, Y.-X. Jiang, J. S. Oh, M. Z. Hasan, K. J. Neubauer, B. Gao, Y. Xie, M. Hashimoto, D. Lu, C. Jozwiak, A. Bostwick, E. Rotenberg, R. J. Birgeneau, J.-H. Chu, M. Yi and P. Dai, \href{https://www.nature.com/articles/s41586-022-05034-z}{Nature \textbf{609}, 490-495 (2022).}

\bibitem{Teng2022_2} X. Teng, J. S. Oh, H. Tan, L. Chen, J. Huang, B. Gao, J.-X. Yin, J.-H. Chu, M. Hashimoto, D. Lu, C. Jozwiak, A. Bostwick, E. Rotenberg, G. E. Granroth, B. Yan, R. J. Birgeneau, P. Dai, M. Yi, \href{https://www.nature.com/articles/s41567-023-01985-w}{Nat. Phys. \textbf{19}, 814 (2023).}

\bibitem{Yin2022} J.-X. Yin, Y.-X. Jiang, X. Teng, Md. S. Hossain, S. Mardanya, T.-R. Chang, Z. Ye, G. Xu, M. M. Denner, T. Neupert, B. Lienhard, H.-B. Deng, C. Setty, Q. Si, G. Chang, Z. Guguchia, B. Gao, N. Shumiya, Q. Zhang, T. A. Cochran, D. Multer, M. Yi, P. Dai, and M. Z. Hasan, \href{https://journals.aps.org/prl/abstract/10.1103/PhysRevLett.129.166401}{Phys. Rev. Lett. \textbf{129}, 166401 (2022).}

\bibitem{Liu2021} F.-H. Liu, W. Fu, Y.-H. Deng, Z.-B. Yuan, and  L.-N. Wu, \href{https://aip.scitation.org/doi/abs/10.1063/5.0053990}{Appl. Phys. Lett. \textbf{119}, 091901 (2021).}

\bibitem{Guster2022} B. Guster, M. Pruneda, P. Ordejón, and E. Canadell, \href{https://journals.aps.org/prb/abstract/10.1103/PhysRevB.105.064107}{Phys. Rev. B. \textbf{105}, 064107 (2022).}

\bibitem{Bjerkelund1964} E. Bjerkelund and A. Kjekshus, \href{https://www.sciencedirect.com/science/article/abs/pii/0022508864900712?via%3Dihub}{J. Less-Common Met. \textbf{7}, 231 (1964).}

\bibitem{Boswell1983} F. W. Boswell, A. Prodan and J. K. Brandon, \href{https://iopscience.iop.org/article/10.1088/0022-3719/16/6/012}{J. Phys. C: Solid State Phys. \textbf{16}, 1067-1076 (1983).}

\bibitem{Mahy1983} J. Mahy, J. Van Landuyt and S. Amelinckx, \href{https://onlinelibrary.wiley.com/doi/abs/10.1002/pssa.2210770151}{Phys. Status Solidi (a) \textbf{77}, K1-K4 (1983).}

\bibitem{Selte1964} K. Selte and A. Kjekshus, \href{http://actachemscand.org/doi/10.3891/acta.chem.scand.18-0690}{Acta Chem. Scand. \textbf{18}, 690-696 (1964).}

\bibitem{Eaglesham1985} D. J. Eaglesham, D. Bird, R. L. Withers and J. W. Steeds, \href{https://iopscience.iop.org/article/10.1088/0022-3719/18/1/008}{J. Phys. C: Solid State Phys. \textbf{18}, 1-11 (1985).}

\bibitem{Morelli1986} R. Morelli, D. Sahu, and M. B. Walker, \href{https://journals.aps.org/prb/abstract/10.1103/PhysRevB.33.4843}{Phys. Rev. B. \textbf{33}, 4843-4848 (1986).}

\bibitem{Boswell1986} F. W. Boswell and A. Prodan, \href{https://journals.aps.org/prb/abstract/10.1103/PhysRevB.34.2979}{Phys. Rev. B. \textbf{34}, 2979(R) (1986).}

\bibitem{Walker1988} M. B. Walker and R. Morelli, \href{https://journals.aps.org/prb/abstract/10.1103/PhysRevB.38.4836}{Phys. Rev. B. \textbf{38}, 4836 (1988).}

\bibitem{Prodan1990} A. Prodan, F. W. Boswell, J. C. Bennett, J. M. Corbett, T. Vidmar, V. Marinkovic and A. Budkowski, \href{https://scripts.iucr.org/cgi-bin/paper?S0108768190004657}{Acta Crystallogr. Sect. B \textbf{46}, 587-591 (1990).}

\bibitem{Bennett1992} J. C. Bennett, S. Ritchie, A. Prodan, F. W. Boswell and J. M. Corbett, \href{https://iopscience.iop.org/article/10.1088/0953-8984/4/9/010}{J. Phys.: Condens. Matter \textbf{4}, 2155 (1992).}

\bibitem{Bronsema1987} K. D. Bronsema, S. V. Smaalen, J. L. De Boer, G. A. Wiegers, F. Jellinek and J. Mahy, \href{https://scripts.iucr.org/cgi-bin/paper?S0108768187097805}{Acta Cryst. B\textbf{43}, 305 (1987).}

\bibitem{Bennett1991} J. C. Bennett, F. W. Boswell, A. Prodan, J. M. Corbett and S. Ritchie, \href{https://iopscience.iop.org/article/10.1088/0953-8984/3/36/001}{J. Phys.: Condens. Matter \textbf{3}, 6959 (1991).}

\bibitem{Kusz1994} J. Kusz and H. B\"{o}hm, \href{https://scripts.iucr.org/cgi-bin/paper?S0108768194005161}{Acta Crystallogr. Sect. B \textbf{50}, 649-655 (1994).}

\bibitem{Kusz1995} J. Kusz, H. B\"{o}hm and J. C. Bennett, \href{https://iopscience.iop.org/article/10.1088/0953-8984/7/14/016}{J. Phys.: Condens. Matter \textbf{7}, 2775-2782 (1995).}

\bibitem{Tadaki1990} S. Tadaki, N. Hino, T. Sambongi, K. Nomura, F. Lévy, \href{https://www.sciencedirect.com/science/article/abs/pii/037967799090107V?via%3Dihub}{Synth. Met. \textbf{38}, 227-234 (1990).}

\bibitem{Zwick1999} F. Zwick, H. Berger, M. Grioni, G. Margaritondo, L. Forró, J. LaVeigne, D. B. Tanner, and M. Onellion, \href{https://journals.aps.org/prb/abstract/10.1103/PhysRevB.59.7762}{Phys. Rev. B. \textbf{59}, 7762 (1999).}

\bibitem{Sambongi1993} T. Sambongi, S. Tadaki, N. Hino, and K. Nomura, \href{https://www.sciencedirect.com/science/article/abs/pii/037967799391122I?via%3Dihub}{Synth. Met. \textbf{58}, 109-114 (1993).}

\bibitem{Luo2017} X. Luo,  F. C. Chen, Q. L. Pei, J. J. Gao, J. Yan, W. J. Lu, P. Tong, Y. Y. Han, W. H. Song, and Y. P. Sun, \href{https://aip.scitation.org/doi/10.1063/1.4977708}{Appl. Phys. Lett. \textbf{110}, 092401 (2017).}

\bibitem{Gao2017} Y. Gao, L. Xu, Y. Qiu, Z. Tian, S. Yuan, and J. Wang, \href{https://aip.scitation.org/doi/10.1063/1.5005907}{J. Appl. Phys. \textbf{122}, 135101 (2017).}

\bibitem{Zhang2020} X. Zhang, Q. Gu, H. Sun, T. Luo, Y. Liu, Y. Chen, Z. Shao, Z. Zhang, S. Li, Y. Sun, Y. Li, X. Li, S. Xue, J. Ge, Y. Xing, R. Comin, Z. Zhu, P.Gao, B. Yan, J. Feng, M. Pan, and J.Wang, \href{https://journals.aps.org/prb/abstract/10.1103/PhysRevB.102.035125}{Phys. Rev. B. \textbf{102}, 035125 (2020).}

\bibitem{Alexandradinata2018} A. Alexandradinata, C. Wang, W. Duan, and L. Glazman, \href{https://journals.aps.org/prx/abstract/10.1103/PhysRevX.8.011027}{Phys. Rev. X. \textbf{8}, 011027 (2018).}

\bibitem{Sun2020} H. Sun, Z. Shao, T. Luo, Q. Gu, Z. Zhang, S. Li, L. Liu, H. Gedeon, X. Zhang, Q. Bian, \href{https://iopscience.iop.org/article/10.1088/1367-2630/aba0657}{New J. Phys. \textbf{22}, 083025 (2020).}

\bibitem{Krause2015} L. Krause, R. Herbst-Irmer, G. M. Sheldrick, and D. Stalke, \href{https://scripts.iucr.org/cgi-bin/paper?S1600576714022985}{J. Appl. Crystallogr. \textbf{48}, 3-10 (2015).}

\bibitem{Sheldrick2015} G. M. Sheldrick, \href{https://scripts.iucr.org/cgi-bin/paper?S2053229614024218}{Acta Crystallogr. Sect. C Struct. Chem. \textbf{71}, 3-8 (2015).}

\bibitem{PBE1996} J. P. Perdew, K. Burke, and M. Ernzerhof, \href{https://journals.aps.org/prl/abstract/10.1103/PhysRevLett.77.3865}{Phys. Rev. B. \textbf{77}, 3865 (1996).}

\bibitem{Kresse1996} G. Kresse and J. Furthmüller, \href{https://journals.aps.org/prb/abstract/10.1103/PhysRevB.54.11169}{Phys. Rev. B. \textbf{54}, 11169 (1996).}

\bibitem{Kresse1996_2} G. Kresse and J. Furthmüller, \href{https://www.sciencedirect.com/science/article/abs/pii/0927025696000080}{Comput. Mater. Sci. \textbf{6}, 15-50 (1996).}

\bibitem{SM} See \href{https://journals.aps.org/prb/supplemental/10.1103/PhysRevB.108.155121}{Supp1emental Material} for divided into two parts (i) physical property characterization and (ii) Electronic band structure. 

\bibitem{Wang2021} V. Wang, N. Xu, J.-C. Liu, G. Tang, W.-T. Geng, \href{https://www.sciencedirect.com/science/article/abs/pii/S0010465521001454}{Comput. Phys. Commun. \textbf{267}, 108033 (2021).}

\bibitem{Sobota2021} J. A. Sobota, Y. He, and Z. -X. Shen, \href{https://journals.aps.org/rmp/abstract/10.1103/RevModPhys.93.025006}{Rev. Mod. Phys. \textbf{93}, 025006 (2021).}

\end{thebibliography}
\end{document}